\documentclass[aps, prd, twocolumn, superscriptaddress, nofootinbib]{revtex4}
\usepackage{graphicx}
\usepackage{dcolumn}
\usepackage{amssymb}
\usepackage{amsmath,amssymb,amsfonts,bbold}

\begin{document}

\newcommand{\Eq}[1]{\mbox{Eq.~(\ref{eqn:#1})}}
\newcommand{\Fig}[1]{\mbox{Fig.~\ref{fig:#1}}}
\newcommand{\Sec}[1]{\mbox{Sec.~\ref{sec:#1}}}

\newcommand{\PHI}{\phi}
\newcommand{\PhiN}{\Phi^{\mathrm{N}}}
\newcommand{\vect}[1]{\mathbf{#1}}
\newcommand{\Del}{\nabla}
\newcommand{\unit}[1]{\;\mathrm{#1}}
\newcommand{\x}{\vect{x}}
\newcommand{\y}{\vect{y}}
\newcommand{\p}{\vect{p}}
\newcommand{\ScS}{\scriptstyle}
\newcommand{\ScScS}{\scriptscriptstyle}
\newcommand{\xplus}[1]{\vect{x}\!\ScScS{+}\!\ScS\vect{#1}}
\newcommand{\xminus}[1]{\vect{x}\!\ScScS{-}\!\ScS\vect{#1}}
\newcommand{\diff}{\mathrm{d}}
\newcommand{\mk}{{\mathbf k}}
\newcommand{\ep}{\epsilon}
\newcommand{\plk}{\mathfrak h}

\newcommand{\be}{\begin{equation}}
\newcommand{\ee}{\end{equation}}
\newcommand{\bea}{\begin{eqnarray}}
\newcommand{\eea}{\end{eqnarray}}
\newcommand{\vu}{{\mathbf u}}
\newcommand{\ve}{{\mathbf e}}
\newcommand{\vn}{{\mathbf n}}
\newcommand{\vk}{{\mathbf k}}
\newcommand{\vz}{{\mathbf z}}
\newcommand{\vx}{{\mathbf x}}
\def\dup{\;\raise1.0pt\hbox{$'$}\hskip-6pt\partial\;}
\def\ddn{\;\overline{\raise1.0pt\hbox{$'$}\hskip-6pt\partial}\;}


\title{The cosmology of minimal varying Lambda theories}
%
%

\author{Stephon Alexander}
\affiliation{Department of Physics, Brown University, Providence, RI, 02906, USA}

\author{Marina Cort\^{e}s}
\affiliation{Perimeter Institute for Theoretical Physics, 31 Caroline Street North,  Waterloo, Ontario N2J 2Y5, Canada}
\affiliation{Instituto de Astrof\'{\i}sica e Ci\^{e}ncias do Espa\c{c}o,
Faculdade de Ci\^{e}ncias, Universidade de Lisboa, 1769-016 Lisboa, Portugal}
\author{Andrew R. Liddle}
\affiliation{Perimeter Institute for Theoretical Physics, 31 Caroline Street North,  Waterloo, Ontario N2J 2Y5, Canada}
\affiliation{Instituto de Astrof\'{\i}sica e Ci\^{e}ncias do Espa\c{c}o,
Faculdade de Ci\^{e}ncias, Universidade de Lisboa, 1769-016 Lisboa, Portugal}
\author{Jo\~{a}o Magueijo}
\affiliation{Theoretical Physics Group, The Blackett Laboratory, Imperial College, Prince Consort Rd., London, SW7 2BZ, United Kingdom}
\author{Robert Sims}
\affiliation{Department of Physics, Brown University, Providence, RI, 02906, USA}
\author{Lee Smolin}
\affiliation{Perimeter Institute for Theoretical Physics, 31 Caroline Street North,  Waterloo, Ontario N2J 2Y5, Canada}

\date{\today}

\begin{abstract}
Inserting a varying Lambda in Einstein's field equations can be made consistent with the Bianchi identities by allowing for torsion, without the need to add scalar field degrees of freedom. In the minimal such theory, Lambda is totally free and undetermined by the field equations in the absence of matter. Inclusion of matter ties Lambda algebraically to it, at least when homogeneity and isotropy are assumed, i.e.\ when there is no Weyl curvature. We show that Lambda is proportional to the matter density, with a proportionality constant  depending on the equation of state. Unfortunately, the proportionality constant becomes infinite for pure radiation, ruling out the minimal theory {\it prima facie} despite of its novel internal consistency. It is possible to generalize the theory still without the addition of kinetic terms, leading to a new algebraically-enforced proportionality between Lambda and the matter density. Lambda and radiation may now coexist in a form consistent with Big Bang Nucleosynthesis, though this places strict constraints on the single free parameter of the theory, $\theta$. In the matter epoch Lambda behaves just like a dark matter component. Its density is proportional to the baryonic and/or dark matter, and its presence and gravitational effects would need to be included in accounting for the necessary dark matter in our Universe. This is a companion paper to Ref.~\cite{us} where the underlying gravitational theory is developed in detail.
\end{abstract}


\maketitle

\section{Introduction}

Allowing for variability of the supposed `constants' of Nature is not without its pitfalls. 
Having assumed their constancy at a foundational level, the formalism often moulds itself to these parameters in ways that lead to contradictions should we try to promote them to dynamical variables. An example is the 
cosmological `constant', which first appeared as a new term in Einstein's equations motivated by the wish for a static Universe:
\be
G_{\mu\nu} + \Lambda g_{\mu\nu}=\kappa \tau_{\mu\nu}\,.
\ee
The Bianchi identities ($\nabla_\mu G^\mu_{\;\nu}=0$) and  metricity condition \mbox{($\nabla_\alpha g_{\mu\nu}=0$)} imply that in vacuum
Lambda must be a constant (and more generally so, if energy--momentum conservation, $\nabla_\mu\tau^\mu_{\;\nu}=0$, is required).   

In a companion paper~\cite{us} we noted that a variable Lambda may be elegantly accommodated by Einstein's
equations should torsion be present. This is most simply implemented in the first-order formalism, either of the Palatini~\cite{textbook} or Pleba\'nski persuasions~\cite{pleb,kirill}.  We first illustrate the construction of this theory with the assumptions
of no matter and vanishing Weyl tensor. Then the Einstein equations reduce to the Self-Dual (SD) condition:\footnote{This is best seen in the Pleba\'nski formalism~\cite{kirill}, where the Einstein equations take the form of the SD condition, plus terms representing the Weyl curvature and the matter content.}
\be\label{SDcond}
R^{ab}=\frac{\Lambda(x)}{3}e^a\wedge e^b\,,
\ee
where $e^a$ are the tetrad fields and $R^{ab}$ is the curvature \mbox{2-form} of the $SO(3,1)$ connection 1-form
$\omega^{ab}$. The Bianchi identities now take the form ${\cal D}R^{ab}=0$, where ${\cal D}$ is the covariant derivative with respect to $\omega^{ab}$. If there is no torsion, defined by $T^a \equiv {\cal D}e^a$, then the Bianchi identities in this context imply $d\Lambda$=0, in agreement with the argument above. In contrast, torsion liberates Lambda from forced constancy in the face of Bianchi identities. Specifically, 
in the context of this illustration (no Weyl, no matter), we would then need: 
\be\label{TLambda0}
T^a=-\frac{1}{2\Lambda} d\Lambda\wedge e^a\,.
\ee
By construction, Lambda would then be left totally undefined in the absence of matter (or Weyl curvature), and should be determined 
by other means. This is not necessarily a drawback, in particular if we have in hand a quantum theory of Lambda predicting its time evolution on purely quantum grounds, and wish to evaluate its implications on classical degrees of freedom via the Einstein equations. 

An action formulation of a theory satisfying these requirements was presented in Ref.~\cite{us}, where it was found that a Gauss--Bonnet (GB) term multiplied by a very specific function of Lambda (and thus no longer a topological term) provided just the required torsion, Eq.~(\ref{TLambda0}), in the absence of Weyl and matter. We can then use this theory to find that in more general situations Lambda, rather than being totally free and unspecified, is algebraically related to the 
matter energy--momentum and the Weyl curvature. The torsion is forced by the theory to take whatever values render the  
consequent variations in Lambda consistent with the Bianchi identities (and a form of matter energy-momentum conservation). In general its value is different from Eq.~(\ref{TLambda0}).

In this paper we examine in detail this theory in the presence of matter, assuming spatial homogeneity and isotropy. We thus preclude a non-vanishing Weyl tensor, an issue that should be borne in mind when extrapolating our conclusions to more general settings. Even with vanishing Weyl curvature we should be aware that the SD condition Eq.~(\ref{SDcond}) has to be abandoned in the presence of matter, because it is inconsistent with the Einstein equations (we revisit the motivating set up of no-matter and no-Weyl in Section~\ref{solsdS}). 
We find that whereas without matter Lambda is totally undetermined, in its presence Lambda becomes tied to the matter, which evolves independently of Lambda, but subjects Lambda to an algebraic constraint involving both the matter energy density and pressure. If  there is no guiding matter, Lambda is completely undetermined, and can be specified arbitrarily. Thus, in this theory Lambda is similar to the cuscuton field~\cite{cuscuton}, which `preys' on matter dynamics, but lacks a dynamics of its own. Unfortunately we find that close to a radiation-dominated epoch, although Lambda would redshift like radiation, it would strongly dominate radiation, leading to a cold Universe. At face value, the theory is therefore ruled out. We examine possible loopholes to this conclusion, none very promising. 

We can relax the condition on the function multiplying the GB term, and leading to torsion given by Eq.~(\ref{TLambda0}) in the absence of Weyl and matter, since this argument is merely motivational. In this paper we allow for this in Section~\ref{general-th} by multiplying the special function of the minimal model by a constant factor $\theta$. 
Then, entirely new behaviour is found for our Friedmann--Robertson--Walker (FRW) reduction. Solutions without matter force Lambda to vanish, instead of allowing it total freedom. Lambda is still tied to matter (should there be matter), but now it cannot exist without matter. In the presence of matter its algebraic dependence on matter density and pressure is different, and dependent on $\theta$, allowing for radiation to coexist with Lambda (with $\Omega_\Lambda$ fixed). Big Bang Nucleosynthesis places strict constraints upon $\theta$, derived here in what should be seen as a preliminary calculation. 

In Section~\ref{concs} we discuss further the implications of our findings.


\section{The basic theory}\label{the-theory}

Let us consider the gravity theory introduced in Ref.~\cite{us}, with matter added on to it. 
For simplicity we assume that the matter action depends only on the matter fields (generically denoted by $\Phi$) 
and the frame fields. We defer to future work the study of matter forms coupled directly to the connection (e.g.\ spinorial matter~\cite{TomK,Toms}). Hence here matter does not {\it directly }generate torsion. However, as we shall see, it still
does so indirectly.

Then, the total action is:
\be
S=\frac{1}{2\kappa}S^{\rm grav}(e,\omega,\Lambda)+S^{\rm M}(\Phi, e)\,,
\ee
where $\kappa=8\pi G$ and
\begin{widetext}
\be\label{action}
S^{\rm grav}= \int_{\cal M}  \epsilon_{abcd} \left ( e^a \wedge e^b \wedge R^{cd}(\omega) 
-\frac{\Lambda}{6}  e^a \wedge e^b \wedge e^c \wedge e^d 
 -\frac{3}{2 \Lambda} 
R^{ab}  \wedge R^{cd} 
\right )\,.
\ee
\end{widetext}
in which $e^a$, $\omega^{ab}$, $\Lambda$, and $\Phi$ are all functions of $x^\mu$. 
Varying with respect to $\omega^{ab}$, $e^a$, and $\Lambda$ generates the equations:
\bea
S^{ab}\equiv T^{[a}\wedge e^{b]}=-\frac{3}{2\Lambda^2}d\Lambda\wedge R^{ab}\label{eq1M}\\
\epsilon_{abcd}{\left(e^b\wedge R^{cd}-\frac{1}{3}\Lambda e^b\wedge e^c\wedge e^d\right)}=-2 \kappa \tau_a \,;\label{eq2M}\\
\epsilon_{abcd}{\left(R^{ab}\wedge R^{cd}-\frac{1}{9}\Lambda^2 e^a \wedge e^b\wedge e^c\wedge e^d\right)}=0 \,,\label{eq3M}
\eea
where
\be
\tau_a =\frac{1}{2}\frac{\delta S^{\rm M}}{\delta e^a} \,,
\ee
is the stress-energy 3-form. Note that from the point of view of the Einstein's equations, Eq.~(\ref{eq2M}), the Lambda
term may be seen as a matter component with:
\be\label{tauLambda}
\tau_a^\Lambda=-\frac{\Lambda}{6\kappa} \epsilon_{abcd} e^b \wedge e^c \wedge e^d \,.
\ee


As explained in the introduction and in Ref.~\cite{us}, if $\tau^a =0$ and the Weyl tensor is zero, these equations are solved by the Self-Dual (SD) condition Eq.~(\ref{SDcond}). Inspection of Eq.~(\ref{eq2M}) shows that the SD condition always solves the Einstein equations {\it in the absence of matter} (although these are not the most general solutions). Then, Lambda is left totally undefined, since the Lambda equation (\ref{eq3M}) produces the tautology $\Lambda^2=\Lambda^2$.
For any (arbitrary) choice of $\Lambda$, Eq.~(\ref{eq1M}) forces the torsion to be Eq.~(\ref{TLambda0}), as required by the Bianchi identities applied to the SD condition. 
We prove below that for a FRW reduction {\it without matter} the SD solutions are indeed the most general ones.

The SD condition is no longer a solution to these equations should there be matter and/or Weyl curvature. 
Specifically, if $\tau^a\neq 0$, then the Einstein Eqs.~(\ref{eq2M}) are no longer 
solved by the SD condition. We will find the solutions below, assuming FRW symmetry. 

Notice that under torsion the stress--energy tensor does not need to be conserved. Applying ${\cal D}$ to the Einstein's equation (\ref{eq2M}), we find that neither the right- or left-hand side have to be zero, with:
\begin{widetext}
\be\label{noncon}
{\cal D}\tau_a = -\frac{1}{2\kappa}  \epsilon_{abcd}\left(T^b\wedge R^{cd}-\Lambda T^b\wedge e^c\wedge e^d-
\frac{d\Lambda}{3}\wedge e^b\wedge e^c\wedge e^d\right) \,.
\ee
\end{widetext}
This is an identity between two 4-forms, so it cannot be contracted with $e^a$. 

However, the following turns out to be true~\cite{textbook}.  Let us break the connection into a torsion-free part, $\tilde \omega^{ab}$ related to the Christoffel connection of $e^a$, and a term encoding the torsion:
\be
\omega^{ab}=\tilde \omega^{ab}(e) +K^{ab} \,.
\ee
$K^{ab}$ is called the contorsion and is related to the torsion by
\be
T^a\equiv De^a=K^a_{\;b}\wedge e^b \,.
\ee
Then, defining $\tilde D$ from $\tilde \omega$, under some circumstances~\cite{textbook} we have that:
\be\label{constfree}
\tilde D\tau_a=0 \,,
\ee
where $\tau_a$ are any of the matter contents derived from $S_M$. This can be proved using Noether's theorem (in combination with the Lie derivative) directly from $S_M$, under certain assumptions (such as that the stress--energy tensor derived from $\tau^a$ is symmetric; see below). In other words, 
matter is covariantly conserved with regards to the torsion-free connection. 

This statement does not apply to Lambda reinterpreted as a matter field, as in Eq.~(\ref{tauLambda}) as we shall see below (and can be seen from the derivation directly by noting that the Lambda term, should Lambda be variable, is not invariant under the relevant Lie derivatives). Indeed the apparent violation of energy conservation due to torsion implied by Eq.~(\ref{noncon}) is there just to cover up for the apparent violations of energy conservation implied by a varying Lambda.

\section{A FRW reduction}
We now assume homogeneity and isotropy. For simplicity we also assume a spatially-flat Universe, but it would be interesting to introduce spatial curvature and 
investigate the implications for the flatness problem. 

\subsection{The geometry}
The FRW ansatz (resulting from homogeneity, parity invariance, and isotropy) is equivalent to the frame fields:
\be
e^0=dt \quad ; \quad
e^i=adx^i \,,
\ee
where $a(t)$ is the expansion factor, $t$ is proper cosmological time and $x^i$ are comoving coordinates.
A general ansatz for the torsion resulting from the symmetries is\footnote{If we allow odd-parity configurations of torsion there is an additional term in the ansatz $T^i = W(t) \epsilon^{ijk} e_j \wedge e_k$.  The phenomenology of this will be developed elsewhere \cite{MZ}.}
\bea
T^0&=&0\label{T0}\\
T^i&=&-T(t)e^0\wedge e^i.\label{Ti}
\eea
Then, $T^a\equiv{\cal D}e^a=de^a+\omega^a_{\;b}\wedge e^b$ 
implies: 
\bea
\omega^i_{\; 0}&=&g(t) e^i =\left(\frac{\dot a}{a}+T\right)e^i\label{omega0i} \,;\\
\omega^i_{\; j}&=&0\,,\label{omegaij}
\eea
and we see that the Hubble parameter in these theories is replaced by: 
\be
g=\frac{\dot a}{a}+T\,.
\ee
We will see in Section~\ref{solsmatt} that typically $g$ is negative for expanding universe solutions.
Therefore, $R^{ab}\equiv d\omega^{ab}+\omega^a_{\;c}\wedge\omega^{cb}$
has  components:
\bea
R^{0i}&=&\frac{1}{a}(ag(t))^.e^0\wedge e^i =\frac{1}{a}(\dot a+Ta)^.e^0\wedge e^i \,;\label{F0i}\\
R^{ij}&=&g^2(t) e^i\wedge e^j= \left(\frac{\dot a}{a}+T\right)^2e^i\wedge e^j \,,\label{Fij}
\eea
where we can recognize the counterparts to the LHS of the  Raychaudhuri and Friedmann equations.
It is easy to check directly that ${\cal D}R=0$ for any function $T(t)$, meaning Bianchi's identities are satisfied. 

\subsection{The matter content}
In general, the stress-energy 3-forms can be related to the tetrad components of the stress-energy tensor, 
$\tau^{ab}$, via:
\be
\tau_a=\frac{1}{6}  \tau_a^{\; b}\epsilon_{bcdf}e^c\wedge e^d\wedge e^f \,.
\ee
Note that the stress--energy tensor $\tau^{ab}$ need not be symmetric if there is torsion~\cite{textbook}. 
However, our symmetries limit the scope of the novelties allowed. As is well known only a perfect fluid is 
allowed under homogeneity and isotropy, with:
\bea
\tau_0&=&-\frac{1}{6}\rho(t)\epsilon_{ijk}e^i \wedge e^j\wedge e^k \,;\\
\tau_i&=&\frac{1}{6}p (t)\epsilon_{ijk}e^0 \wedge e^j\wedge e^k\,.
\eea
Hence, $\tau^{ab}$ must be symmetric (and indeed diagonal), satisfying the conditions leading to Eq.~(\ref{constfree}).
As one can see, the cosmological constant  reinterpreted as a matter content (c.f.\ Eq.~\ref{tauLambda}) leads to $\rho=-p=\Lambda/\kappa$. Thus, its variation would naively seem to imply violations of energy conservation. This is where torsion comes in, as we shall see below. 


\subsection{The field equations}
Inserting these expressions into the field equations produces
\begin{eqnarray}
T&=&\frac{3\dot\Lambda}{2\Lambda^2}g^2\label{T1CM}\,;\\
g^2=\left (\frac{\dot a}{a}+T\right)^2&=&\frac{\Lambda+\kappa\rho }{3}\label{F1CM}\,;\\
g^2+2\frac{(ag)^.}{a}&=&-\kappa p+\Lambda\label{F2CM0}\,;\\
g^2\frac{1}{a}(ag)^.&=&\frac{\Lambda^2}{9}\label{LeqCM0} \,,
\end{eqnarray}
for Eq.~(\ref{eq1M}), Eq.~(\ref{eq2M}) with index $a=0$ and $a=i$, and Eq.~(\ref{eq3M}), respectively.

These can be re-arranged as:
\begin{widetext}
\bea
T&=&\frac{\dot\Lambda}{2\Lambda}\left(1+\frac{\kappa\rho}{\Lambda}\right)\label{T1CMb} \,; \\
g^2=\left (\frac{\dot a}{a}+\frac{\dot\Lambda}{2\Lambda}\left(1+\frac{\kappa\rho}{\Lambda}\right)\right)^2&=&\frac{\Lambda+\kappa\rho }{3}\,;\\
\frac{(ag)^.}{a}=\frac{1}{a}\left(\dot a+\frac{\dot\Lambda}{2\Lambda}a\left(1+\frac{\kappa\rho}{\Lambda}\right)\right)^.&=& \frac{\Lambda}{3}-\frac{\kappa}{6}(\rho +3p) \,; \label{F2CM}\\
(\Lambda+\kappa\rho )
\left(\Lambda-\frac{\kappa}{2}(\rho +3p)\right)&=&\Lambda^2\,.\label{LeqCM}
\eea
\end{widetext}
Indeed by combining Eqs.~(\ref{T1CM}) and  (\ref{F1CM}) we can obtain an explicit expression for the torsion, as in Eq.~(\ref{T1CMb}), and this inserted back into Eq.~(\ref{F1CM}) produces a version of the Friedman equation.
Also, the Friedman equation (\ref{F1CM}) can be combined with Eq.~(\ref{F2CM0}) to produce our version of the
Raychaudhuri equation, Eq.~(\ref{F2CM}). Finally the Friedman and Raychaudhuri equations can be inserted into
Eq.~(\ref{LeqCM0}) to write the Lambda equation as an algebraic constraint, Eq.~(\ref{LeqCM}). 

These are the central equations of this theory. 

\subsection{The conservation equation}\label{cons}

Finally, in lieu of the Raychaudhuri equation it may be useful to use the conservation equation for matter. 
In view of the comments leading to Eq.~(\ref{constfree}) we know that it must be the case that:
\be\label{nonconF2c}
\dot\rho+3\frac{\dot a}{a}(\rho+p)=0 \,.
\ee
This is a really interesting result. As we see, matter is covariantly conserved with respect to 
the torsion-free connection. The torsion is present only to account for the non-conservation of the Lambda energy--momentum. To see this fact, it is interesting to carry out some strictly-speaking unnecessary algebra, to see how 
Eq.~(\ref{nonconF2c}) emerges in the face of the apparent violations of energy conservation due to torsion (c.f.\ Eq.~(\ref{noncon})).


Let us specialize the non-conservation equation (\ref{noncon})  for the FRW case, leading to:
\be\label{nonconF1}
\dot\rho+3\frac{\dot a}{a}(\rho+p)+3Tp=\frac{1}{\kappa}\left(-3g^2 T+3\Lambda T- \dot\Lambda\right)\,. 
\ee
Alternatively we could have obtained this expression by multiplying Eq.~(\ref{F1CM}) by $a^2$ and dotting it,  and comparing with Eq.~(\ref{F2CM}),
resulting in:
\be\label{nonconF2b}
\dot\rho+3\frac{\dot a}{a}(\rho+p)=-T(\rho+3p)+\frac{2\Lambda T-\dot\Lambda}{\kappa} \,.
\ee
Using Eq.~(\ref{F1CM}) we can see that the two expressions are equivalent. 

Upon substituting the expression for $T$ (i.e.\ Eq.~\ref{T1CMb}) and using Eq.~(\ref{LeqCM}) we do get
Eq.~(\ref{nonconF2c}). Why this mysterious cancellation? 
Equations (\ref{F1CM})  and  (\ref{F2CM}) can be written in a neat and suggestive way as:
\bea
g^2&=&\frac{\kappa\bar \rho}{3}\label{F1CMc} \,;\\
\frac{(ag)^.}{a}&=&-\frac{\kappa}{6}(\bar\rho+3\bar p)\,,\label{F2CMc}
\eea
where we have repackaged Lambda and matter as a two-component fluid:
\bea
\bar\rho&=&\rho+\rho_\Lambda \,;\\
\bar p&=&p +p_\Lambda \,,
\eea
with $p_\Lambda=-\rho_\Lambda=-\Lambda/\kappa$.
Multiplying the first (Friedmann) equation by $a^2$ and only then dotting, and using the second equation, then leads to:
\be\label{barrhoc}
\dot{\bar\rho}+3\frac{\dot a}{a}(\bar\rho+\bar p)=-T(\bar\rho+3\bar p) \,.
\ee
This is equivalent to:
\bea
\dot\rho+3\frac{\dot a}{a}(\rho+p)&=&0 \,;\\
\dot\rho_\Lambda +3\frac{\dot a}{a}(\rho_\Lambda+p_\Lambda)&=&\frac{\dot\Lambda}{\kappa} =-T(\bar\rho+3\bar p)\label{Lambdaeq3} \,,
\eea
since the last equation 
is just another way to state the the RHS of Eq.~(\ref{nonconF2b}) is zero, as we have checked. 
Physically it means that the apparent violations of energy conservation implied by Eq.~(\ref{barrhoc}) all go towards covering up the apparent violations of energy conservation implied by a varying $\Lambda$ in the face of
Eq.~(\ref{Lambdaeq3}).


\section{De Sitter-like solutions}\label{solsdS}

We now examine FRW solutions with Lambda but no matter. We already know that the SD condition solves the equations; here we prove that they are the most general solutions with FRW symmetry. Setting $\rho=p=0$ in Eqs.~(\ref{T1CMb})--(\ref{LeqCM}), gives:
\bea
T(t)&=&\frac{\dot\Lambda}{2\Lambda}\label{Toft}\,;\\
g^2=\left(\frac{\dot a}{a}+\frac{\dot\Lambda}{2\Lambda}\right)^2&=&\frac{\Lambda}{3}\label{FdS}\,;\\
\frac{(ag)^.}{a}=\frac{1}{a}\left(\dot a+\frac{\dot\Lambda}{2\Lambda}a\right)^.&=&\frac{\Lambda}{3}\label{RdS}\,;\\
\Lambda^2&=&\Lambda^2\label{Taut}\,.
\eea
Considering Eqs.~(\ref{F0i}) and  (\ref{Fij}), we see that Eqs.~(\ref{FdS}) and (\ref{RdS}) are nothing but the 
SD condition Eq.~(\ref{SDcond}), which therefore is the most general FRW solution without matter.
Concomitantly, Eq.~(\ref{Toft}) is just an expression of Eq.~(\ref{TLambda0}) for FRW, while  Eq.~(\ref{Taut}) is  
the tautology already highlighted after Eq.~\ref{SDcond}. Due to this tautology $\Lambda$ is left undetermined.

It is easy to see that dotting Eq.~(\ref{FdS})  gives Eq.~(\ref{RdS}). This might provide a simple
solution to the tension between Friedman and  Raychaudhuri equations often found in causal set arguments 
for a randomly-evolving Lambda. If $\Lambda$ is a constant this reduces to the usual equations with zero torsion and constant Lambda. 

The general solution of the theory is therefore an arbitrarily chosen $\Lambda(t)$, for which the torsion
is obtained via Eq.~(\ref{Toft}). Setting $b=\sqrt\Lambda a$ in Eq.~(\ref{FdS}) leads to
\be
\left(\frac{\dot b}{b} \right)^2=\frac{\Lambda}{3} \,,
\ee
which can be integrated, leading to expansion factor:
\be
a=a_0\frac{e^{\int\frac{\Lambda(t)}{3}dt}}{\sqrt{\Lambda(t)}} \,. 
\ee
These expressions represent generalizations of the cosmological patch of de Sitter space-time, allowing for arbitrary time-varying Lambda. It would be interesting to obtain the full manifold. Also, there might be SD solutions beyond these ones, i.e.\ SD solutions which do not have a FRW representation.

\section{FRW solutions with matter}\label{solsmatt}

Replacing Eq.~(\ref{F2CM}) by Eq.~(\ref{nonconF2c}), one obtains the following useful complete set of equations for this theory in the presence of matter:
\bea
\left (\frac{\dot a}{a}+T \right)^2&=&\frac{\Lambda+\kappa\rho }{3}\label{eq1} \,;\\
T&=&\frac{\dot\Lambda}{2\Lambda}\left(1+\frac{\kappa\rho}{\Lambda}\right)\label{eq2} \,;\\
\dot\rho+3\frac{\dot a}{a}(\rho+p)&=&0\label{eq3} \,;\\
(\Lambda+\kappa\rho )
\left(\Lambda-\frac{\kappa}{2}(\rho +3p)\right)&=&\Lambda^2\,.\label{eq4}
\eea
They are
composed, respectively, of the Friedman equation including torsion, the torsion equation, the matter equation and the Lambda equation.  Examining the Lambda equation we see that,
whereas $\Lambda$ is left undefined if there is no matter, in contrast, should there be matter, Lambda is forced to track it 
as an algebraic constraint. Indeed, Eq.~(\ref{eq4}) implies:
\be\label{ratio1}
\Lambda=\kappa\rho\frac{1+3w}{1-3w} \,,
\ee
or 
\be
\Omega_\Lambda\equiv \frac{\rho_\Lambda}{\rho+\rho_\Lambda}=\frac{1+3w}{2} \,,
\ee
(with $\rho_\Lambda=\Lambda/\kappa$). We note the singular cases:
\begin{itemize}
\item For radiation, $w=1/3$, we are forced by the constraint to have $\rho=0$, so the only solutions are those of Section~\ref{solsdS}. 
\item For $w>1/3$, $\Lambda<0$ and $|\Lambda|>\rho$, so there are no solutions (the RHS of Eq.~(\ref{eq1}) must be positive).
\item For $w=-1/3$ we have $\Lambda\rho=0$, so in addition to the solutions of Section~\ref{solsdS} we can set $\Lambda=0$ and have Milne ($a\propto t$).
\end{itemize}
In the other cases we have solutions for which $\Lambda\propto \rho$, with $\Lambda>0$ for $-1/3<w<1/3$, and $\Lambda<0$ for $w<-1/3$. If $w$ is constant,  Eqs.~(\ref{ratio1}) and  (\ref{eq3}) imply
 \be\label{scale3}
\Lambda\propto\rho\propto  \frac{1}{ a^{3(1+w)}} \,,
\ee
or equivalently:
\be
\frac{\dot\Lambda}{\Lambda}=\frac{\dot\rho}{\rho}=-3(1+w)\frac{\dot a}{a} \,.
\ee
Hence, the torsion is proportional to the Hubble parameter, with:
\be
T=\frac{\dot\Lambda}{\Lambda}\frac{1}{1+3w}=-3\frac{1+w}{1+3w}\frac{\dot a}{a} \,,
\ee
where we have used Eqs.~(\ref{eq2}) and (\ref{ratio1}) for the first identity. The effect of torsion and Lambda upon the Friedman equation (\ref{eq1}) is therefore simply to renormalize the gravitational constant. Indeed, Eq.~(\ref{eq1}) can be written as:
\bea
\left (\frac{\dot a}{a}\right)^2&=&\frac{\bar\kappa\rho }{3}\\
\bar \kappa&=&\frac{\kappa}{2}\frac{(1+3w)^2}{1-3w}\,.
\eea
As usual
\be
a\propto t^{2/[3(1+w)]}\,,
\ee
but now the relation between the Hubble parameter and the density feels a renormalized gravitational constant, with 
the following implications.

\subsection{Matter and radiation epochs}
Specifically, for matter ($w=0$) we have the standard 
\bea
a&\propto& t^{2/3}\\
 \Lambda&\propto&\rho \propto 1/a^3 \,,
\eea
but now torsion and Lambda work to renormalize $G$ by a factor of 1/2. Therefore, this has the opposite effect of ``Lambda as dark matter''. This is a bit counterintuitive, since  the constraint forces $\Lambda=\kappa \rho$, so one might think that Lambda would increase $G$. However the torsion in the LHS of the Friedman equation acts to reduce the effect upon the Hubble parameter. 

Radiation is a singular case, as we have seen, but
even if we were to look at radiation as a $w\rightarrow 1/3^-$ limit, this would be disastrous.
Although Lambda would then track radiation, with
\bea
 \Lambda&\propto&\rho \propto 1/a^4\,;\\
a&\propto& t^{1/2}  \,,
\eea 
the point is that the proportionality constant between Hubble and the temperature would be infinite. This would conflict seriously with Big Bang Nucleosynthesis (BBN). It would have the same effect as a very large number of relativistic degrees of freedom $g_\star$, or a very large $G$ as in Brans--Dicke theories. 

\subsection{A matter and radiation fluid}\label{radwithmat}
Adding the subdominant matter to the radiation does not help to solve the situation in the radiation epoch.
Let
\bea
\rho&=&\rho_r+\rho_m \,;\\
p&=&p_r=\frac{1}{3}\rho_r \,,
\eea
and examine the $\rho_r\gg \rho_m$ regime. Then, Eq.~(\ref{eq4}) implies:
\bea
\Lambda & = & \kappa \frac{2 \rho_r ^2+ \rho_m ^2+3\rho_m \rho_r}{\rho_m}\nonumber\\
&\approx& 2\kappa \rho_r\frac{\rho_r}{\rho_m}\propto\frac{1}{a^5} \,,
\eea
with $\Lambda\gg \rho_r$. Hence 
\be
T\approx \frac{\dot \Lambda}{2\Lambda}\approx -\frac{5}{2}\frac{\dot a}{a} \,,
\ee
from which we infer
\be
\frac{9}{4}\left(\frac{\dot a}{a}\right)^2\approx\frac{\Lambda}{3}\propto \frac{1}{a^5} \,. 
\ee
The joint system in the `radiation' epoch therefore behaves like a fluid with $w=2/3$ and has
\be
a\propto t^{2/5} \,.
\ee

\subsection{Accelerating solutions}\label{acc}

Our theory cannot use Lambda to generate the current acceleration of the Universe. 
As a curiosity we investigate how Lambda behaves in this theory if some other source (quintessence \cite{Tsujikawa:2013fta} or an inflaton field)
causes accelerated expansion. We find that in our theory Lambda would still `track'  a matter source with $w=-1$,
thereby behaving like a conventional Lambda in that case. The constraints derived above translate into: 
\bea
\Lambda&=&-\frac{\kappa \rho}{2} \,;\\
\bar \kappa&=&\frac{\kappa}{2} \,,
\eea
that is, Lambda would acquire the opposite sign of what is causing the acceleration, reducing its effective Hubble constant.  

\section{More general theories}\label{general-th}

Among the many arguments leading to the action Eq.~(\ref{action}) was the motivation of allowing a varying Lambda in the specific case of no matter and no Weyl curvature, in the face of Bianchi identities, as explained in the Introduction. This fixes the very specific 
 form of the 
pre-factor,  $h$, of the Gauss--Bonnet term:
\be\label{hcrit}
h=-\frac{3}{2\Lambda} \,. 
\ee
This choice leads to the above homogeneous and isotropic cosmology, where Lambda is totally free without matter, but otherwise must track matter algebraically. Torsion, rather than being given by Eq.~(\ref{TLambda0}), is generally of the form Eq.~(\ref{T1CMb}).

We may, however, consider more general functions $h$, such as:
\be\label{hgen}
h=-\frac{3\theta}{2\Lambda} \,,
\ee
with $\theta$ a constant. This would generalize our model even without the addition of kinetic terms 
(which would bring it close to a more conventional quintessence model~\cite{Tsujikawa:2013fta}).
Then, the Lambda equation (\ref{eq3M}) would become:
\bea
\epsilon_{abcd}{\left(R^{ab}\wedge R^{cd}-\frac{1}{6 h'} e^a \wedge e^b\wedge e^c\wedge e^d\right)}&=&
\nonumber\\
\epsilon_{abcd}{\left(R^{ab}\wedge R^{cd}-\frac{\Lambda^2}{9 \theta} e^a \wedge e^b\wedge e^c\wedge e^d\right)}&=&
0 \,.
\eea
The Einstein equation (\ref{eq2M}) is unmodified, but the torsion equation (\ref{eq1M}) now reads:
\bea
S^{ab}\equiv T^{[a}\wedge e^{b]}&=&-h'd\Lambda\wedge R^{ab}\nonumber \\
&=&-\frac{3\theta}{2\Lambda^2}d\Lambda\wedge R^{ab} \,.
\eea
Following the calculations described above for the FRW reduction we can obtain the set of equations:
\bea
\left (\frac{\dot a}{a}+T \right)^2&=&\frac{\Lambda+\kappa\rho }{3}\label{Neq1} \,;\\
T&=&\frac{\theta\dot\Lambda}{2\Lambda}\left(1+\frac{\kappa\rho}{\Lambda}\right)\label{Neq2} \,;\\
\dot\rho+3\frac{\dot a}{a}(\rho+p)&=&0\label{Neq3} \,;\\
\theta(\Lambda+\kappa\rho )
\left(\Lambda-\frac{\kappa}{2}(\rho +3p)\right)&=&\Lambda^2\,.\label{Neq4}
\eea
As before, matter is covariantly conserved with respect to the torsion-free connection. 
This can be checked directly (with some tedious algebra) mimicking the calculation of Section~\ref{cons}, or by appealing to the general argument following Eq.~(\ref{constfree}) in Section~\ref{the-theory}. 

In contrast with the choice Eq.~(\ref{hcrit}), there cannot be any FRW solutions without matter and $\Lambda\neq 0$, if $\theta\neq 1$.
Setting $\rho=p=0$ in Eq.~(\ref{Neq4}) we obtain:
\be
\Lambda^2=\theta\Lambda^2 \,,
\ee
so that either $\theta=1$ and we have the situation described in Section~\ref{solsdS} (Lambda is undetermined), or $\theta\neq 1$,
in which case the only solution is $\Lambda=0$ (i.e.\ Minkowski space-time). 

If there is matter, on the other hand, we can start by solving Eq.~(\ref{Neq4}), with solutions: 
\be
r=\frac{\kappa\rho}{\Lambda}=\frac{1}{2}\left[ \frac{1-3w}{1+3w} \pm \sqrt{\left(\frac{1-3w}{1+3w}\right)^2
+\frac{8(\theta-1)}{\theta(1+3w)}}\label{root1}
\right] \,.
\ee
The $\theta = 1$ results are recovered by taking the positive root for $w \geq -1/3$ and the negative one for $w <-1/3$.
We find that provided $\theta>1$, radiation ($w=1/3$) may now coexist with Lambda, bypassing what seems a strong no-go
situation for the minimal model ($\theta=1$). We can follow through the 
equivalent of the calculations in Section~\ref{solsmatt} to find that the radiation epoch 
proceeds in the same way as in the standard Hot Big Bang cosmology, but with a renormalized
gravitational constant:
\be\label{kapparen}
\bar\kappa=\kappa\frac{1+\sqrt\frac{\theta}{\theta-1}}{\left[1-2\theta\left(1+\sqrt{\frac{\theta-1}{\theta}}\right)\right]^2} \,.
\ee
We see that the effective gravitational constant is unchanged if and only if $\theta$ takes on the special value $9/8$. In this case, each side of Eq.~(\ref{Neq1}) is enhanced by a factor 4, due to the torsion on the left-hand side (which again dominates, flipping the sign of the expansion term) and to the presence of $\Lambda$ (which acts as dark radiation with with three times the normal radiation density) on the right-hand side. 

Any other value of $\theta$ leads to a renormalization of $G$ during the radiation era, which is strongly constrained by nucleosynthesis. The usual BBN constraint~\cite{uzan,Olive}  is given by:
\be
-0.10<\frac{\Delta G}{G}<0.13 \,,
\ee
and translates into a range of width only 0.015 about the special value of $\theta = 9/8$, i.e.\
\be
1.11<\theta<1.14 \,.
\ee
Note that, unlike in Section~\ref{radwithmat}, 
given that the radiation solution has a non-vanishing
ratio $\kappa\rho/\Lambda$, the presence of a sub-dominant dust component in the radiation epoch has negligible
effect.

We stress that the renormalization of $G$ encoded in Eq.~(\ref{kapparen}) is an algebraic relation with respect to the $G$ measured 
in a Cavendish balance experiment.\footnote{With the proviso that the counterpart to the Schwarzchild solution in this theory does not force a further
renormalization of a truly bare $G$ in the vicinity of the Sun.} It is not a cumulative change to be tracked throughout the history of the universe 
(as in Brans--Dicke theories), so we do not need to incorporate the matter and late-time accelerating epoch renormalizations into 
the BBN constraint.

On the other hand observations made in the matter and late-time acceleration epochs should take into account that the Universe
behaves as usual in these theories, but also with a renormalized $G$, given by:
\be\label{kapparen1}
\bar\kappa=\kappa\frac{1+1/r}{\left[1-\frac{3\theta}{2}(1+w)(1+r)\right]^2} \,,
\ee
(with $r$ given by Eq.~(\ref{root1})). This formula generalizes Eq.~(\ref{kapparen}), and should be used to compute the renormalized $G$
when  $w=0$ and  $w\approx -1$. Once $\theta$ is fixed (e.g.\ by the BBN) there is no further leverage in the predictions 
for the matter and  accelerating epochs. For the matter era with $\theta = 9/8$ the gravitational constant, $G$ is reduced by a factor of about 4; the $\Lambda$ density is 0.84 of the matter density but the torsion increases the effective ${(\dot a}/a)^2$ on the left-hand side of Friedmann in Eq.~(\ref{Neq1}) by a factor of 7:
\be
\frac{{\dot g}}{g}= -2.69\frac{{\dot a}}{a} \,.
\ee
Lambda may be seen as a dark matter component, but notice that due to the effects of torsion its gravitational effect is different from what would have been expected from the energy density in Lambda alone.  

In Eq.~(\ref{root1}) we picked the root that continuously goes to the result obtained for $\theta=1$, but this is not a watertight argument. Both roots solve the equations and each should be assessed for physical relevance. One might also ask whether the system could or should switch roots when the combined fluid content evolves across $w = -1/3$, as is likely to be required of a viable cosmological model. Such investigations are beyond the scope of the present paper.

\section{Conclusions}\label{concs}

It is undoubtedly disappointing that a theory with {\it fewer} parameters than General Relativity (as argued in Ref.~\cite{us})
should be felled at the most undemanding first obstacle.  The adage ``one should never let observations get in the way of a good idea" 
might be invoked here, were it not for the non-existence of a radiation epoch presenting such a damning verdict on any theory claiming even a remote
connection with reality. We have tried -- and failed -- to find a way around this result. This does not mean that it does not exist. 
In the absence of a solution to such a glaring contradiction with reality we have to content ourselves with cataloguing this theory
as a ``good idea on paper, that did not fulfil its promise in the face of reality''.\footnote{We remark that
the steady-state Universe, whether a good idea or not, suffers from very similar observational problems.} 
Unless, of course, we have got it all wrong with basic Big Bang cosmology.

The introduction of a new parameter, $\theta$, places the theory on the same footing as General Relativity regarding the
number of free parameters. With this addition the theory does not crash out at the first hurdle, but that does not mean it will
lead to a viable cosmology. If it does, that would be by itself a remarkable result. We stress that beside $\theta$ there are no further free parameters in the theory, unlike in {\it models} designed to fit reality at any cost~\cite{Tsujikawa:2013fta}.  
In Section~\ref{general-th} we
have done some preliminary work on the constraints BBN places upon $\theta$.
If $\theta$ is indeed a constant this fixes the predictions made for all other observational tests we have to 
satisfy.

We find that in the narrow band allowed by BBN for $\theta$, in the matter epoch Lambda behaves like a dark matter component with a gravitational effect equivalent to reducing by roughly 4 times that of the baryonic and other dark matter components. This factor is different from what would  have been inferred solely from Lambda's  energy density because in this theory Lambda induces torsion, and this has a gravitational effect upon expansion of its own. 
The above is true at zeroth order, i.e.\ ignoring cosmological perturbations, and how they feed into some of the constraints.
The calculation of all observational fits should be redone in the context of a perturbed Universe, for the self-consistency of our matter epoch constraints. The presence of Lambda and its gravitational effects will have to be included in the inventory of the necessary dark matter in our Universe.

Obviously we would still have to explain the current acceleration of our Universe by other means. Whatever they might be, Lambda would play a role and interact with them, as we have already described in Section~\ref{acc}. 
The hope remains that an adaptation of our theory would not only keep Lambda at bay throughout the life of the Universe, but explain the late-time acceleration we appear to have observed. In this respect one should not forget that the classical model proposed in Ref.~\cite{us} had its roots in considerations on the quantum nature of the cosmological constant \cite{paper0,Lambda-quant}. Perhaps this radical step is unavoidable for a proper understanding of the cosmological constant and the current acceleration of the Universe.

\begin{acknowledgments}
We thank Ted Jacobson, Kirill Krasnov, Roger Penrose,  Syksy R\"as\"anen, and Tom Zlo\'snik for discussions on matters related to this paper. 
This research was supported in part by Perimeter Institute for Theoretical Physics. Research at Perimeter Institute is supported by the Government of Canada through Industry Canada and by the Province of Ontario through the Ministry of Research and Innovation. This research was also partly supported by grants from NSERC and FQXi. 
M.C.\ was supported by Funda\c{c}\~{a}o para a Ci\^{e}ncia e a Tecnologia (FCT) through grant SFRH/BPD/111010/2015 (Portugal). J.M.\ was supported by STFC consolidated grant number ST/L00044X/1. M.C., A.R.L., and L.S.\ are especially thankful to the John Templeton Foundation for their generous support of this project. 
\end{acknowledgments}



\begin{thebibliography}{99}
\bibitem{us}
S. Alexander, M. Cort\^{e}s, A. R. Liddle, J. Magueijo, R. Sims, and L. Smolin,
{\it A Zero-Parameter Extension of General Relativity with Varying Cosmological
Constant}, simultaneously submitted, arXiv:1905.10380 [gr-qc].

\bibitem{textbook}
M. Gockeler and T. Schucker, {\it Differential Geometry, Gauge Theories, and Gravity (Cambridge Monographs on Mathematical Physics)}, CUP, Cambridge, 1989. 

\bibitem{pleb}
J. F. Pleba\'nski, 
{\it On the separation of Einsteinian substructures},
J. Math. Phys. {\bf 18}, 2511
(1977).

\bibitem{kirill}
  K. Krasnov,
  {\it Pleba\'nski Formulation of General Relativity: A Practical Introduction},
  Gen.\ Rel.\ Grav.\  {\bf 43}, 1 (2011),
  [arXiv:0904.0423 [gr-qc]].

\bibitem{cuscuton} 
  N. Afshordi, D. J. H. Chung, and G. Geshnizjani,
  {\it Cuscuton: A Causal Field Theory with an Infinite Speed of Sound},
  Phys.\ Rev.\ D {\bf 75}, 083513 (2007)
  [hep-th/0609150].

\bibitem{Toms} 
  J. Magueijo, T. G. Zlo\'snik, and T. W. B. Kibble,
  {\it Cosmology with a spin},
  Phys.\ Rev.\ D {\bf 87}, 063504 (2013),
  [arXiv:1212.0585 [astro-ph.CO]].

\bibitem{TomK}
  T. W. B. Kibble,
  {\it Lorentz invariance and the gravitational field},
  J.\ Math.\ Phys.\  {\bf 2}, 212 (1961),

\bibitem{MZ} J. Magueijo and T. G. Zlo\'snik, {\it Parity violating Friedmann Universes},  arXiv:1908.05184 [gr-qc].

\bibitem{uzan}
  J. P. Uzan,
  {\it Varying Constants, Gravitation and Cosmology},
  Living Rev.\ Rel.\  {\bf 14}, 2 (2011),
  [arXiv:1009.5514 [astro-ph.CO]].

\bibitem{Olive} 
  R. H. Cyburt, B. D. Fields, K. A. Olive, and E. Skillman,
  {\it New BBN limits on physics beyond the standard model from $^4He$},
  Astropart.\ Phys.\  {\bf 23}, 313 (2005),
  [astro-ph/0408033].

\bibitem{Tsujikawa:2013fta} 
  S. Tsujikawa,
  {\it Quintessence: A Review},
  Class.\ Quant.\ Grav.\  {\bf 30}, 214003 (2013)
  [arXiv:1304.1961 [gr-qc]].

\bibitem{paper0}
  S. Alexander, J. Magueijo, and L. Smolin,
  {\it The quantum cosmological constant},
  Symmetry {\bf 11}, 1130 (2019) [arXiv:1807.01381 [gr-qc]].

\bibitem{Lambda-quant}
  J. Magueijo and L. Smolin,
  {\it A Universe that does not know the time},
Universe {\bf 5}, 84 (2019)
  [arXiv:1807.01520 [gr-qc]].
  
\end{thebibliography}
\end{document}